\documentclass[11pt]{article}
\usepackage{graphicx}
\usepackage{enumerate}

\usepackage{booktabs}
\usepackage{amsmath, amssymb, amsfonts}
\usepackage[mathcal]{euscript}
\usepackage{tabularray}
\UseTblrLibrary{amsmath}
\usepackage{multirow}
  
\newtheorem{lemma}{Lemma}

\usepackage[commandnameprefix=ifneeded]{changes} 
\usepackage{geometry}  
\usepackage{todonotes}

\usepackage{graphicx}
\usepackage{xcolor}
\usepackage{tikz}
\usepackage{scalerel}

\usepackage{ctable}      
\usepackage{caption}
\usepackage{lscape}    
\usepackage{threeparttable} 

\usepackage[authoryear]{natbib}
\usepackage{hyperref}
\hypersetup{colorlinks,allcolors=blue}

\usepackage{url} 

\newcommand{\blind}{1}

\begin{document}
\def\spacingset#1{\renewcommand{\baselinestretch}{#1}\small\normalsize} \spacingset{1}

\if1\blind
{
  \title{\bf Estimating the Heritability of Longitudinal Rate-of-Change: Genetic Insights into PSA Velocity in Prostate Cancer-Free Individuals}
  \author{%
    Pei Zhang \thanks{The authors gratefully acknowledge support from the NIH Intramural Research Program} \hspace{.2cm} \\ 
    Department of Mathematics, University of Maryland, College Park  \\
    Division of Cancer Epidemiology and Genetics, National Cancer Institute, MD, USA \\ [1.5em]
    Xiaoyu Wang \\
    Division of Cancer Epidemiology and Genetics, National Cancer Institute, MD, USA \\[1.5em]
    Jianxin Shi \thanks{Correspondence: \href{mailto: jianxin.shi@nih.gov}{jianxin.shi@nih.gov}}
     \\
    Division of Cancer Epidemiology and Genetics, National Cancer Institute, MD, USA \\[1.5em]
    Paul S. Albert \thanks{Correspondence: \href{mailto: albertp@nih.gov}{albertp@nih.gov} } \\
    Division of Cancer Epidemiology and Genetics, National Cancer Institute, MD, USA \\
  }
  \date{}     
  \maketitle
 } \fi

 \if0\blind
 {
  \bigskip
  \bigskip
  \bigskip
  \begin{center}
   {\LARGE\bf Estimating the Heritability of Longitudinal Rate-of-Change: Genetic Insights into PSA Velocity in Prostate Cancer-Free Individuals}
 \end{center}
  \medskip
 } \fi

\bigskip
\begin{abstract}   

Serum prostate-specific antigen (PSA) is widely used for prostate cancer screening. While the genetics of PSA levels has been studied to enhance screening accuracy, the genetic basis of PSA velocity, the rate of PSA change over time, remains unclear. The Prostate, Lung, Colorectal, and Ovarian (PLCO) Cancer Screening Trial, a large, randomized study with longitudinal PSA data (15,260 cancer-free males, averaging 5.34 samples per subject) and genome-wide genotype data, provides a unique opportunity to estimate PSA velocity heritability. We developed a mixed model to jointly estimate heritability of PSA levels at age 54 and PSA velocity. To accommodate the large dataset, we implemented two efficient computational approaches: a partitioning and meta-analysis strategy using average information restricted maximum likelihood (AI-REML), and a fast restricted Haseman-Elston (REHE) regression method. Simulations showed that both methods yield unbiased estimates of both heritability metrics, with AI-REML providing smaller variability in the estimation of velocity heritability than REHE. Applying AI-REML to PLCO data, we estimated heritability at 0.32 (s.e. = 0.07) for baseline PSA and 0.45 (s.e. = 0.18) for PSA velocity. These findings reveal a substantial genetic contribution to PSA velocity, supporting future genome-wide studies to identify variants affecting PSA dynamics and improve PSA-based screening.

\end{abstract}

\noindent%

{\it Keywords:} linear mixed model; AI-REML algorithm; REHE method; meta-analysis; truncation
\vfill

\newpage
\spacingset{1.9} 
\section{Introduction} \label{sec:intro}
Serum prostate-specific antigen (PSA) measurement is the most widely used screening test for prostate cancer \citep{catalona1991measurement, adhyam2012review}. There has been extensive work in evaluating the clinical utility of PSA for the early detection of prostate cancer \citep{catalona1991measurement, adhyam2012review}. However, the effectiveness of screening strategies using PSA remains controversial. Recent approaches that have shown some promise involve incorporating  genetics to improve screening efficiency \citep{kachuri2023genetically, zhang2025mixed}. Characterizing the heritability of PSA among individuals who do not develop prostate cancer is important, as it reflects the proportion of PSA variation attributable to genetic factors in a natural population. Most studies have estimated this heritability from a cross-sectional perspective. However, PSA is a biomarker that varies over time and strongly associated with age. Therefore, it is important to distinguish between the heritability at a specific age (or time point) and the heritability of rate of change in this biomarker across time. In this article, we develop novel approaches for simultaneously estimating these two components of heritability.

\begin{figure}[!ht]
\centering
\includegraphics[width= 10cm]{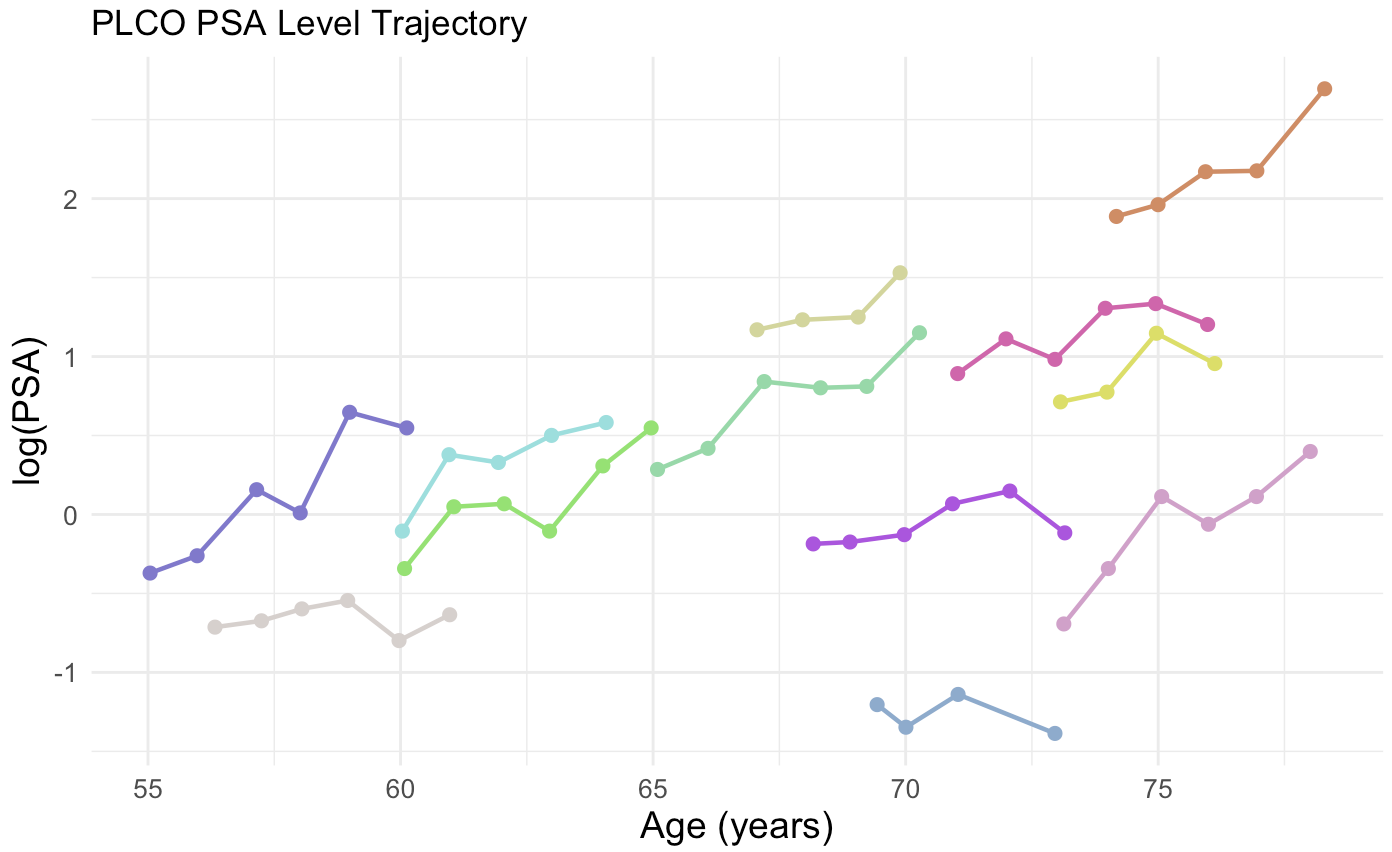}
\caption{PSA Level Trajectories of 12 Randomly Selected Participants from the 15{,}260 Healthy European White Males in the PLCO Cancer Screening Trial}
\label{fig:figure_PSA_Log_50_application}%
\end{figure}

Even among individuals who are not diagnosed with prostate cancer, there is marked variation in both the overall level and the rate-of-change in PSA across individuals. Figure~\ref{fig:figure_PSA_Log_50_application} shows serial annual PSA measurements for 12 randomly selected males from the screening arm of the Prostate, Lung, Colorectal, and Ovarian (PLCO) Cancer Screening Trial who were not diagnosed with prostate cancer. This figure shows marked variation in both the initial PSA measurement as well as the rate of change of PSA over time. What fraction of the variation can be attributable to genome-wide common variants? Is the fraction for the rate-of-change different than that for a cross-sectional assessment at a particular age?  New methodological approaches are needed to address these questions, even though \citet{zhang2025mixed} have examined these two components attributable to statistically significant variants associated with PSA.

Estimating heritability is one of central problems in human genetics and statistical methods have been developed based on twin studies and other family designs. Recently, many statistical methods have been proposed to estimate heritability explained by genome-wide common variants using individual-level genotype and phenotype data \citep{yang2010common,  lango2010hundreds, lippert2011fast, yang2011gcta, speed2014multiblup, chen2014estimating} or summary statistics of genome-wide association studies \citep{bulik2015ld, finucane2015partitioning, speed2020evaluating, li2023accurate}. However, these methods  incorporate only individual-level genetic effects on the intercept and are not suitable for estimating heritability components in both the static and dynamic aspects of longitudinal phenotypes.  \cite{van2004genetic} developed a two-stage modeling approach for estimating those heritability components in twin studies where the genetic structure can be inferred. In the current article,  we use genome-wide variants to determine the genetic structure needed for estimating these metrics in population-based studies.

The PLCO Cancer Screening Trial is a randomized controlled trial designed to evaluate various screening strategies for early cancer detection \citep{gohagan2000prostate, pinsky2012prostate, machiela2023gwas}. This trial has collected comprehensive data, including repeated measurements of serum PSA concentrations and genome-wide genotypic information based on genotyping arrays. For this study, we focused on a subsample of $15,260$ male participants from the PLCO study who had no history of cancer or prostate-related surgeries and had at least one recorded PSA measurement. This data enabled us to estimate the heritability metrics on both the baseline PSA levels and PSA velocity in a natural population.

In this article, we develop a linear mixed model to jointly estimate the heritability of PSA at both a single time point (or age) and the velocity of the trajectory across time.  Several challenges arise in this estimation, including variation from ultra-high-dimensional SNP-specific random effects (nearly 7 million common variants) and implementation in population-based studies with in large sample sizes. We develop two estimation procedures with one being likelihood-based and the other moment-based focused on minimizing the sum of squared differences of products of all paired measurements. The likelihood-based approach is potentially very efficient but suffers from computational infeasibility when the total number of measurements is large. Alternatively, the method-of-moments approach is more computationally feasible, but suffers from reduced efficiency.  With regard to the likelihood-based approach, we propose to first partition subjects into non-overlapping groups, analyze data in each group, and develop a novel meta-analytic approach that makes estimation possible even in large studies.
This is compared with the method-of-moments approach in a wide range of situations through simulations as well as using analysis from the PLCO study. 

The rest of this article is organized as follow. In Section~\ref{section2_2}, we first briefly review the methods for estimating heritability for quantitative traits based on cross-sectional studies and typical method for estimating the heritability using individual-level genotype and phenotype data, and then formulate a mixed modeling framework to jointly estimate heritability of PSA level at a a particular age and PSA velocity using longitudinal observations. We then developed methods for estimating the model using the {AI-REML} algorithm and REHE regression method.  Section~\ref{section3_2} assesses the performance of the two approaches using simulations. Section~\ref{section4_2} estimates variance components, heritability of PSA baseline and velocity, and their 95\% confidence intervals based on the longitudinal PSA levels in the PLCO study. In Section~\ref{section5_2}, we offer further discussion on the findings and implications.

\section{Methodologies}  \label{section2_2}
\subsection{Estimating Heritability in Cross-Sectional Studies}
For a quantitative trait, heritability quantifies the proportion of variation in a trait that can be attributed to genetic differences among individuals. Two types of heritability are often used in the genetics field: 
\(H^2\), the broad sense heritability that measures the full contribution of genetics to the trait, and \(h^2\), the narrow sense heritability that measures the total additive genetic effects. The analysis of narrow sense heritability can be based on the following model
\begin{equation} \label{Eq:model1}
    y_{i} = \mu + g_{i} + \epsilon_{i}, 
\end{equation}
where $y_i$ denotes the phenotypic value, \(\mu\) denotes the mean of phenotypic values,  $g_i = \sum_{p=1}^{P}\alpha_p z_{ip}$ denotes the total additive genetic effects, $z_{ip}$ represents the genotypic values standardized with mean zero and variance one across subjects, $\alpha_p \sim \mathcal{N}(0, \sigma^2_{\alpha})$ is random effect for SNP \(p\), and  $\epsilon_i$ denotes the residue. Under the assumption that $g_i$ and $\epsilon_i$ are independent, we have  $\text{Var}(y_i)=\sigma_g^2+\sigma_\epsilon^2$ with $\sigma_g^2= \text{Var}(g_i) = P \sigma^2_{\alpha}$ and $\sigma_\epsilon^2= \text{Var}(\epsilon_i)$. The narrow sense heritability can be defined as $h^2=\sigma^2_{g}/{(\sigma^2_{g}+\sigma^2_{\epsilon})}$ \citep{visscher2008heritability}.

Estimating \(h^2\) relies on the following equation \(cor(y_i, y_k) = h^2 G_{ik}\) after we standardize \(\boldsymbol{y} = (y_1, y_2, \ldots, y_N)^{\top}\)such that the variance of the phenotype is 1, where \(G_{ik} = \sum_{p=1}^{P} z_{ip}z_{kp}/P\) is the genetic correlation between two subjects. This equation leads to two types of methods for estimating \(h^2\): 
restricted maximum likelihood (REML) \citep{yang2010common} and phenotype correlation–genotype correlation (PCGC)\citep{lynch1998genetics, golan2014measuring}. Since we will extend both methods for estimating variance components in longitudinal settings, we here briefly describe the two methods.  
The PCGC method regresses the empirical phenotypic correlation \(y_i y_k\) onto genetic correlations  \(G_{ik}\) with the slope unbiasedly estimating  \(h^2\) \citep{lynch1998genetics, golan2014measuring}, similar to the Haseman-Elston (HE) regression in the genetic literature \citep{haseman1972investigation, chen2014estimating}. This method has been further extended to estimate multiple variance components and is referred to as the Restricted Haseman-Elston (REHE) method \citep{yue2021rehe}. The confidence interval of the estimate is obtained by parametric bootstrapping. This approach, while simple and easy to implement, loses efficiency compared to those methods based on the full likelihood.   

We then review the REML method.  Let \(\boldsymbol{Z}= (z_{ip})_{N \times P}\) denote the  genetic information matrix for \(N\) subjects and \(P\) common variants. Let \(\boldsymbol{G} = \boldsymbol{Z}\boldsymbol{Z}^{\top}/P\) be the \(N \times N\) genetic correlation matrix. Let \(\boldsymbol{\theta} = (\sigma^2_g, \sigma^2_{\epsilon})^{\top}\).  The marginal distribution \(\boldsymbol{y}| \boldsymbol{Z},  \boldsymbol{\theta}, \boldsymbol{\beta} \sim \mathcal{N}(\boldsymbol{A\boldsymbol{\beta}}, \boldsymbol{V})\),  where \(\boldsymbol{A}\) and \(\boldsymbol{\beta}\) are the design matrix and coefficients for fixed effects, respectively,  \( \boldsymbol{V} = \text{Cov}(\boldsymbol{y} | \boldsymbol{Z}, \boldsymbol{\theta}) = \sigma^2_g \boldsymbol{G}  + \sigma^2_{\epsilon} \boldsymbol{I}_{N}\) and \(\boldsymbol{I}_{N}\) is a \(N \times N\) identity matrix.  The restricted log likelihood can be expressed as
\begin{equation*}
   \ell_{REML}(\boldsymbol{\theta} | \boldsymbol{y}, \boldsymbol{Z}) 
    = -\frac{1}{2}\left\{ \boldsymbol{y}^{\top}\boldsymbol{R}\boldsymbol{y} + \text{log} \ \text{det}(\boldsymbol{V}) +  \text{log} \ \text{det}(\boldsymbol{A}^{\top}\boldsymbol{V}^{-1}\boldsymbol{A})\right\} + \text{constant},
\end{equation*}
where \(\boldsymbol{R} = \boldsymbol{V}^{-1} - \boldsymbol{V}^{-1} \boldsymbol{A}(\boldsymbol{A}^{\top}\boldsymbol{V}^{-1}\boldsymbol{A})^{-1}\boldsymbol{A}^{\top}\boldsymbol{V}^{-1}\) represents the matrix transforming observed response into residuals.  The average information REML (AI-REML) algorithm \citep{johnson1995restricted,gilmour1995average} is easier to implement, computationally efficient, and numerically stable to unbiasedly estimate variance components and has been widely used for fitting mixed model, including estimating heritability in genetics \citep{yang2010common} . In each iteration, we calculate the gradient vector \(\boldsymbol{DL}_{\boldsymbol{\theta}}(\boldsymbol{\theta})\) with  
\begin{equation} \label{Eq:DI}
    \frac{\partial \ell_{REML}}{\partial \theta_{s}} =  -\frac{1}{2} \left\{ \text{tr}\left\{ \boldsymbol{R}\frac{\partial \boldsymbol{V}}{\partial \theta_{s}}\right\} -  \boldsymbol{y}^{\top}\boldsymbol{R}\frac{\partial \boldsymbol{V}}{\partial \theta_{s}}\boldsymbol{R} \boldsymbol{y}  \right\},
\end{equation}
and \(\boldsymbol{AI}_{\boldsymbol{\theta}}(\boldsymbol{\theta})\),  the average of the observed and expected information matrix of \(\boldsymbol{\theta}\), with 
\begin{equation} \label{Eq:AI}
    AI_{\boldsymbol{\theta}}( \boldsymbol{\theta})_{s,k}   = \frac{1}{2}( \frac{\partial^{2}\ell_{REML}}{\partial \theta_{s} \theta_{k}} + \mathbb{E}[\frac{\partial^{2}\ell_{REML}}{\partial \theta_{s} \theta_{k}}]  ) 
    = -\frac{1}{2} \left\{ \boldsymbol{y}^{\top}\boldsymbol{R}\frac{\partial \boldsymbol{V}}{\partial \theta_{s}} \boldsymbol{R}\frac{\partial \boldsymbol{V}}{\partial \theta_{k}} \boldsymbol{R} \boldsymbol{y} \right\},
\end{equation}
where \(\partial{\boldsymbol{V}}/ \partial{\theta_1} =  \boldsymbol{G}\) and \(\partial{\boldsymbol{V}}/ \partial{\theta_2} =  \boldsymbol{I}_N\). Of note, the average information matrix is much easier to calculate.  The AI-REML method updates \(\boldsymbol{\theta}\) at iteration \(h+1\) as: 
\begin{equation} \label{Eq:iteration}
   \hat{\boldsymbol{\theta}}^{(h+1)} =  \hat{\boldsymbol{\theta}}^{(h)}  - [\boldsymbol{AI}_{\boldsymbol{\theta}}(\hat{\boldsymbol{\theta}}^{(h)})]^{-1} \boldsymbol{DL}_{\boldsymbol{\theta}}(\hat{\boldsymbol{\theta}}^{(h)}),
\end{equation}
where \(h\) is iterate number and each \(\theta_{s}\) and \(\theta_{k}\) are the \(s\)th and \(k\)th element in \(\boldsymbol{\theta}\). 

The covariance matrix for \(\hat{\boldsymbol{\theta}} \) is obtained as  \(-\boldsymbol{AI}_{\boldsymbol{\theta}}(\hat{\boldsymbol{\theta}} )^{-1}\) after convergence and the variance of  \( \hat{h}^2\) can be derived by the Delta method \citep{lynch1998genetics, lehmann2006theory}. Of note, \(\boldsymbol{V}\) can be expressed as a linear combination of multiple variance components, a necessary condition for the AI-REML algorithm \citep{johnson1995restricted}.

\subsection{A Mixed Modeling Approach for Estimating Intercept Heritability and Velocity Heritability in Longitudinal Studies}
Consider a longitudinal study with \(N\) subjects, each of which has \(n_i\) observations. For the \(j\)th screening for subject \(i\), let \(y_{ij}\) be the quantitative trait value, \(t_{ij}\) be the time variable, and \(m_{ij}^{(l)}\)  be the value of covariate \(l (l=1,\ldots,L)\) and \(m_{ij}^{(1)} = t_{ij} \). The study has \(P\) common variants with genotypic values \(z_{ip}\) standardized to have mean zero and variance one across subjects. We propose the following mixed effect model:
\begin{equation} \label{Eq:model2}
    y_{ij} = \beta_{0}  + \sum_{l=1}^{L}\beta_{l} m_{ij}^{(l)} + g_{i} + g_{i}^{\ast} t_{ij} + b_{i0} + b_{i1}t_{ij} + e_{ij},
\end{equation}
where \( b_{i0} \sim \mathcal{N}(0, \sigma^2_{b0}) \) is subject-specific random effects on the intercept, \( b_{i1} \sim \mathcal{N}(0, \sigma^2_{b1}) \) is the random effect on the slope,  \(g_{i} = \sum_{p=1}^{P} \alpha_{p}z_{ip}\) denotes baseline random genetic effects, \(g_{i}^{\ast} = \sum_{p=1}^{P}\eta_{p}z_{ip}\) denotes the random effects for the slope, and residual \( e_{ij} \sim \mathcal{N}(0, \sigma^2_{e}) \).   We assume that genotype random effects \(\alpha_{p} \sim \mathcal{N}(0, \sigma^2_{\alpha})\) and \(\eta_{p} \sim \mathcal (0, \sigma^2_{\eta})\), and they are mutually independent, and are independent of the residuals. We have \(\sigma^2_{g}= \text{Var}(g_i) = P\sigma^2_{\alpha}\) and \(\sigma^2_{g^{\ast}} = \text{Var}(g^{\ast}_i)= P\sigma^2_{\eta} \). The heritability of phenotypic trait at the intercept is defined as  
\begin{equation} \label{Eq: baseline_heritability}
     \lambda_{1} = \sigma^2_{g}/(\sigma^2_{g} + \sigma^2_{b_{0}}), 
\end{equation}
and the heritability of the phenotypic trait's velocity is defined as 
\begin{equation} \label{Eq: slope_heritability}
\lambda_{2} = \sigma^2_{g^{\ast}}/(\sigma^2_{g^{\ast}} + \sigma^2_{b_{1}}).
\end{equation}

The definition of $\lambda_1$ requires us to specify a particular time point of interest. As expressed in Equation~\eqref{Eq:model2}, this corresponds to the case where the time variable is set to zero ($t_{ij}=0$). The metric $\lambda_1$ is different from what is usually assessed for heritability at a particular time point in cross-sectional studies. In a cross-sectional design, we cannot distinguish between individual-specific and residual variation, resulting in heritability defined as $\sigma^2_{g}/(\sigma^2_{g} + \sigma^2_{b0}+\sigma^2_e)$ in our notation.  Our definition for $\lambda_1$ uses the longitudinal data structure to assess the heritability at a particular time point treating the within-individual variation as measurement error.

\subsubsection{Leveraging AI-REML Algorithm for Studies with Small-to-Moderate Sample Sizes}
Let $\boldsymbol{\theta}= (\sigma^2_{g}, \sigma^2_{g^{\ast}}, \sigma^2_{b0}, \sigma^2_{b1}, \sigma^2_e)^{\top}$.  We now estimate \(\boldsymbol{\theta}\) by maximizing the restricted log likelihood function. To proceed, let \(\boldsymbol{y} = (\boldsymbol{y}_1^{\top}, \ldots, \boldsymbol{y}_{N}^{\top})^{\top}\) represent the vector of all observed phenotypic traits, where \(\boldsymbol{y}_i = (y_{i1}, \ldots, y_{in_i})^{\top}\) represents the traits for the \(i\)th subject.  The marginal 
distribution for  \(\boldsymbol{y}\) is given as 
\begin{equation} \label{Eq7:matrix}
    \boldsymbol{y} | \boldsymbol{Z}, \boldsymbol{\theta}, \boldsymbol{\beta} \sim \mathcal{N}(\boldsymbol{A}\boldsymbol{\beta}, \boldsymbol{V}),
\end{equation}
where \(\boldsymbol{V} = \text{Cov}(\boldsymbol{y} | \boldsymbol{Z}, \boldsymbol{\theta})\) is the \(\sum_{i=1}^{N} n_i \times \sum_{i=1}^{N} n_i \) covariance matrix of \(\boldsymbol{y}\),  \(\boldsymbol{A}\) is the \( \sum_{i=1}^{N} n_i \times (L+1)\) covariate matrix and \(\boldsymbol{\beta} = (\beta_0, \ldots, \beta_{L})^{\top}\) denotes the \( (L+1) \times 1\) column vector of coefficients for fixed effects.  Then, the restricted log likelihood can be expressed as 
\begin{equation} \label{Eq7:REML}
    \ell_{REML}(\boldsymbol{\theta} | \boldsymbol{y}, \boldsymbol{Z}) 
    = -\frac{1}{2}\left\{ \boldsymbol{y}^{\top}\boldsymbol{R}\boldsymbol{y} + \text{log} \ \text{det}(\boldsymbol{V}) +  \text{log} \ \text{det}(\boldsymbol{A}^{\top}\boldsymbol{V}^{-1}\boldsymbol{A})\right\} + \text{constant},
\end{equation}
where \(\boldsymbol{R} = \boldsymbol{V}^{-1} - \boldsymbol{V}^{-1} \boldsymbol{A}(\boldsymbol{A}^{\top}\boldsymbol{V}^{-1}\boldsymbol{A})^{-1}\boldsymbol{A}^{\top}\boldsymbol{V}^{-1}\) represents the matrix transforming observed response into residuals. 

To use the AI-REML algorithm to estimate variance components $\boldsymbol{\theta}$, it is necessary to show that the covariance matrix $\boldsymbol{V}$ can be expressed as the linear combination of all variance components in \(\boldsymbol{\theta}\) (Lemma~\ref{lem:V}). 

\begin{lemma}   \label{lem:V}
In Model~\eqref{Eq:model2}, the covariance matrix \(\boldsymbol{V}\) is a linear combination of variance components, denoted as \( \boldsymbol{V} = \sum_{s=1}^{5} \theta_{s} \boldsymbol{H}_{s} \), where
\begin{equation*}
\begin{split}
        & \boldsymbol{H}_{1} = \begin{pmatrix}  \boldsymbol{1}_{n_{1}} \boldsymbol{1}_{n_{1}}^{\top}  G_{11} &   \boldsymbol{1}_{n_{1}} \boldsymbol{1}_{n_{2}}^{\top} G_{12} &   \cdots &  \boldsymbol{1}_{n_{1}} \boldsymbol{1}_{n_{N}}^{\top}G_{1N}  \\
        \boldsymbol{1}_{n_{2}} \boldsymbol{1}_{n_{1}}^{\top}G_{21} &  \boldsymbol{1}_{n_{2}} \boldsymbol{1}_{n_{2}}^{\top} G_{22} &   \cdots &  \boldsymbol{1}_{n_{2}} \boldsymbol{1}_{n_{N}}^{\top} G_{2N} \\
        \vdots & \vdots & \ddots & \vdots \\
        \boldsymbol{1}_{n_{N}} \boldsymbol{1}_{n_{1}}^{\top}G_{N1} &  \boldsymbol{1}_{n_{N}} \boldsymbol{1}_{n_{2}}^{\top} G_{N2} &   \cdots &  \boldsymbol{1}_{n_{N}} \boldsymbol{1}_{n_{N}}^{\top} G_{NN} \\
        \end{pmatrix},    \\
        & \boldsymbol{H}_{2} = \begin{pmatrix}  \boldsymbol{t}_{1} \boldsymbol{t}_{1}^{\top}G_{11} &  \boldsymbol{t}_{1} \boldsymbol{t}_{2}^{\top}G_{12} &   \cdots &  \boldsymbol{t}_{1} \boldsymbol{t}_{N}^{\top}G_{1N}  \\
        \boldsymbol{t}_{2} \boldsymbol{t}_{1}^{\top}G_{21} &  \boldsymbol{t}_{2} \boldsymbol{t}_{2}^{\top} G_{22} &   \cdots &  \boldsymbol{t}_{2} \boldsymbol{t}_{N}^{\top} G_{2N} \\
        \vdots & \vdots & \ddots & \vdots \\
       \boldsymbol{t}_{N} \boldsymbol{t}_{1}^{\top}G_{N1} &  \boldsymbol{t}_{N} \boldsymbol{t}_{2}^{\top} G_{N2} & \cdots &  \boldsymbol{t}_{N} \boldsymbol{t}_{N}^{\top}G_{NN}  \\
        \end{pmatrix},    \\
        & \boldsymbol{H}_{3} =  \begin{pmatrix}
            \boldsymbol{J}_{n_{1}} & \boldsymbol{0} &  \cdots & \boldsymbol{0} \\
            \boldsymbol{0} & \boldsymbol{J}_{n_{2}} & \cdots & \boldsymbol{0} \\
            \vdots & \vdots  & \ddots & \vdots  \\
           \boldsymbol{0}  & \cdots &  \boldsymbol{0}&   \boldsymbol{J}_{n_{N}}
        \end{pmatrix},   \boldsymbol{H}_{4} =   \begin{pmatrix}
            \boldsymbol{t}_{1}\boldsymbol{t}_{1}^{\top} &  \boldsymbol{0} &  \cdots & \boldsymbol{0} \\
            \boldsymbol{0} &  \boldsymbol{t}_{2}\boldsymbol{t}_{2}^{\top}& \cdots & \boldsymbol{0} \\
           \vdots & \vdots  & \ddots & \vdots  \\
            \boldsymbol{0}  & \cdots &  \boldsymbol{0} &  \boldsymbol{t}_{N}\boldsymbol{t}_{N}^{\top}
        \end{pmatrix},  \boldsymbol{H}_{5} =  \boldsymbol{I}_{\sum_{i=1}^{N}n_{i}},
    \end{split}
\end{equation*}
\(\boldsymbol{G}= \boldsymbol{Z}\boldsymbol{Z}^{\top}/P\) is the \(N \times N \) genetic relationship matrix, \(G_{ik}\) is the \((i,k)\)th element of \(\boldsymbol{G}\) and \(\boldsymbol{I}_{s}\) is a \( s \times s \) identity matrix, \(\boldsymbol{J}_{s}\) is a \( s \times s \) matrix with every element equal to \(1\), for any positive integer number \(s\).
\end{lemma}

The proof of Lemma~\ref{lem:V} is provided as Supplementary Material A. 
Using the notation introduced in Lemma~\ref{lem:V}, we derive the gradient function of the restricted log likelihood function 
\begin{equation*}
\boldsymbol{DL}_{\boldsymbol{\theta}}(\boldsymbol{\theta}) = \frac{ \partial \ell_{REML}}{ \partial \boldsymbol{\theta}}= -\frac{1}{2} \begin{pmatrix}
        \text{tr}(\boldsymbol{R}\boldsymbol{H}_{1}) - \boldsymbol{y}^{\top}\boldsymbol{R}\boldsymbol{H}_{1}\boldsymbol{R}\boldsymbol{y} \\ 
        \text{tr}(\boldsymbol{R}\boldsymbol{H}_{2}) - \boldsymbol{y}^{\top}\boldsymbol{R}\boldsymbol{H}_{2}\boldsymbol{R}\boldsymbol{y} \\ 
         \vdots   \\
         \text{tr}(\boldsymbol{R}\boldsymbol{H}_{5}) - \boldsymbol{y}^{\top}\boldsymbol{R}\boldsymbol{H}_{5}\boldsymbol{R}\boldsymbol{y} 
    \end{pmatrix},
\end{equation*}
\begin{equation*}
\boldsymbol{AI}_{\boldsymbol{\theta}}(\boldsymbol{\theta}) = -\frac{1}{2} \begin{pmatrix}
\boldsymbol{y}^{\top}\boldsymbol{R}\boldsymbol{H}_{1}\boldsymbol{R}\boldsymbol{H}_{1}\boldsymbol{R}\boldsymbol{y} &  \boldsymbol{y}^{\top}\boldsymbol{R}\boldsymbol{H}_{1}\boldsymbol{R}\boldsymbol{H}_{2}\boldsymbol{R}\boldsymbol{y}  & \cdots & \boldsymbol{y}^{\top}\boldsymbol{R}\boldsymbol{H}_{1}\boldsymbol{R}\boldsymbol{H}_{6}\boldsymbol{R}\boldsymbol{y}  \\
\boldsymbol{y}^{\top}\boldsymbol{R}\boldsymbol{H}_{2}\boldsymbol{R}\boldsymbol{H}_{1}\boldsymbol{R}\boldsymbol{y} &  \boldsymbol{y}^{\top}\boldsymbol{R}\boldsymbol{H}_{2}\boldsymbol{R}\boldsymbol{H}_{2}\boldsymbol{R}\boldsymbol{y}  & \cdots & \boldsymbol{y}^{\top}\boldsymbol{R}\boldsymbol{H}_{2}\boldsymbol{R}\boldsymbol{H}_{6}\boldsymbol{R}\boldsymbol{y}  \\ 
    \vdots & \vdots & \ddots & \vdots \\ 
\boldsymbol{y}^{\top}\boldsymbol{R}\boldsymbol{H}_{5}\boldsymbol{R}\boldsymbol{H}_{1}\boldsymbol{R}\boldsymbol{y} &  \boldsymbol{y}^{\top}\boldsymbol{R}\boldsymbol{H}_{5}\boldsymbol{R}\boldsymbol{H}_{2}\boldsymbol{R}\boldsymbol{y}  & \cdots & \boldsymbol{y}^{\top}\boldsymbol{R}\boldsymbol{H}_{5}\boldsymbol{R}\boldsymbol{H}_{5}\boldsymbol{R}\boldsymbol{y}  \\
\end{pmatrix}.
\end{equation*}
In the AI-REML algorithm, we iterate $\hat{\boldsymbol{\theta}}^{(h+1)} =  \hat{\boldsymbol{\theta}}^{(h)}  - [\boldsymbol{AI}_{\boldsymbol{\theta}}(\hat{\boldsymbol{\theta}}^{(h)})]^{-1} \boldsymbol{DL}_{\boldsymbol{\theta}}(\hat{\boldsymbol{\theta}}^{(h)})$ until convergence with \( |\ell_{REML}(\hat{\boldsymbol{\theta}}^{(h+1)}) - \ell_{REML} (\hat{\boldsymbol{\theta}}^{(h)}) | < 10^{-4}\). Of note, in the iteration process, if one parameter or multiple parameters are updated to be negative, we set it or them to be \( \sigma^2_{ph} \times 10^{-6}\) \citep{yang2011gcta} where \(\sigma^2_{ph}\) denotes the empirical variance of all observed phenotypic values.  Let \(\hat{\boldsymbol{\theta}}\) be the REML estimate. The baseline heritability is estimated as  $\hat{{\lambda}}_{1} = \hat{\sigma}^2_{g}/ ( \hat{\sigma}^2_{g}  + \hat{\sigma}^2_{b0})$ and the velocity heritability is estimated as  $ \hat{\lambda}_{2} =  \hat{\sigma}^2_{{g}^{\ast}}/ (\hat{\sigma}^2_{{g}^{\ast}} + \hat{\sigma}^2_{b1})$.

We use the Delta method to obtain the covariance matrix  \(\text{Cov}( (\hat{\lambda}_1, \hat{\lambda}_2)^{\top})\). To proceed, we define \( \boldsymbol{\xi} =(\xi_1, \xi_2, \xi_3, \xi_4,\xi_5)^{\top} := (\lambda_{1}, \lambda_{2}, \sigma^2_{g} + \sigma^2_{b0}, \sigma^2_{g^{*}} + \sigma^2_{b1}, \sigma^2_{e})^{\top} \). In Supplementary Material B, we derive the Jacobian matrix
\begin{equation*}
    \boldsymbol{K} = (\frac{\partial \boldsymbol{\theta}}{ \partial \boldsymbol{\xi}}) = \begin{pmatrix}
        \xi_3 & 0 & \xi_1 & 0 & 0 \\
        0 & \xi_4 & 0 & \xi_2 & 0 \\
        -\xi_3 & 0 & 1-\xi_1 & 0 & 0 \\
        0 & -\xi_4 & 0 & 1-\xi_2 & 0 \\
        0 & 0 & 0 & 0 & 1
    \end{pmatrix}.
\end{equation*}
By the Delta method \citep{lynch1998genetics, lehmann2006theory},  the estimated Fisher's information matrix for \(\boldsymbol{\xi}\) evaluated at \(\hat{\boldsymbol{\xi}}\) is derived as 
\begin{equation} \label{Eq: AI_xi}
-\boldsymbol{AI}_{\boldsymbol{\xi}}(\boldsymbol{\hat{\xi}}) \approx -\hat{\boldsymbol{K}}^{\top} \boldsymbol{AI}_{\boldsymbol{\theta}}(\hat{\boldsymbol{\theta}})\hat{\boldsymbol{K}},
\end{equation} 
where \(\boldsymbol{AI}(\hat{\boldsymbol{\theta}})\) is the average information matrix for \(\boldsymbol{\theta}\) evaluated at \(\hat{\boldsymbol{\theta}}\). The covariance matrix for \(( \hat{\lambda}_1, \hat{\lambda}_2)^{\top}\) can be extracted from the inverse of \( -  \boldsymbol{AI}_{\boldsymbol{\xi}}(\boldsymbol{\hat{\xi}}) \).

\subsubsection{Leveraging AI-REML Algorithm by Partitioning and Meta-analysis for Studies with Large Sample Sizes} 
Of note, the AI-REML algorithm requires inverting the covariance matrix \(\boldsymbol{V}\), which is a  \(\sum_{i=1}^{N} n_i \times \sum_{i=1}^{N} n_i\) dimensional matrix. This becomes computationally prohibitive when \( \sum_{i=1}^{N} n_i\) is very large. An approach to address the challenge is to evenly divide the subjects into 
\(M\) non-overlapping groups and then apply the AI-REML algorithm for each group to estimate \( \hat{\boldsymbol{\theta}}_{m} ( \text{and} \ \hat{\boldsymbol{\xi}}_{m} )\).  The final estimate can then be obtained by appropriately combining \( \hat{\boldsymbol{\theta}}_{m} (  \text{and} \ \hat{\boldsymbol{\xi}}_{m} )\), using methods such as a simple average or a fixed effect meta-analysis. 
While this strategy is conceptually straightforward, we have two comments regarding its implementation. First, this approach may lose efficiency because it does not utilize cross-group information. To minimize information loss, we recommend reducing the number of groups as much as possible while maintaining computational feasibility. Second, all elements of \(\boldsymbol{\theta}\) and \(\boldsymbol{\xi} \) are constrained to be positive. Failing to account for this left truncation could result in biased estimates for \(\boldsymbol{\theta}\) and \(\boldsymbol{\xi} \), as we will show in simulations. In what follows, we propose a likelihood-based approach that accounts for truncation to estimate each parameter, its variance, and 95\% confidence interval.

Assume \(X_m = \mu + \epsilon_m \) for $m=1,\cdots, M$, where \(\epsilon_m \sim \mathcal{N}(0, \sigma^2_m)\) and \(\sigma^2_m \) is known. 
Given an observation \(X_m=x_m\) left-truncated at zero, the likelihood is 
\(f_m(x_m) = (1- \pi_m)\delta_0(x_m) + \frac{1}{\sigma_m}\phi ( \frac{x_m- \mu}{\sigma_m}) \). 
Here, \(\pi_m = P(X_m > 0| X_m \sim \mathcal{N}(\mu, \sigma^2_m) \text{\ without\ truncation})= \Phi(\frac{ \mu}{\sigma_m}) \), \(\delta_0(x)\) is the Dirac delta function at zero, \(\phi(x)\) is the probability density function  of standard Gaussian distribution \( \mathcal{N}(0,1)\), and $\Phi(x)$ is the cumulative distribution function of \(\mathcal{N}(0,1)\).  Then, the logarithmic likelihood for the observed data \(X_1, \ldots, X_M\) is derived as 
\[
\ell(\mu | X_1, \ldots, X_M)  = \sum_{ \{m : X_m=0 \}} \log  \{ 1- \Phi(\frac{\mu}{\sigma_m}) \}  + \sum_{\{m: X_m >0\}} \{ - \log \sigma_m + \log \phi (\frac{X_m - \mu}{\sigma_m})\}. 
\]
Then, we obtain the maximum likelihood estimator (MLE) \(\hat{\mu}\) and its variance. In Supplementary Material D, we present simulations comparing the performance of three approaches: the simple averaging method, the fixed-effect meta-analysis, and our proposed meta-analysis method that accounts for left truncation. Additionally, in Supplementary Material D, we provide MLE for estimating \(\mu\) when observations are doubly truncated - left-truncated at zero and right-truncated at one - which is applicable for estimating \( \lambda_1\) and \( \lambda_2\).

\subsubsection{Estimating Heritability Metrics Using the Restricted Haseman-Elston Regression Method} \label{section2.3.2}

\cite{yue2021rehe} proposed a method-of-moments-based estimation procedure, termed restricted Haseman-Elston (REHE) regression  Method, to ensure non-negative variance component estimates for cross-sectional traits, thereby providing heritability estimates constrained between 0 and 1. Building on this approach, we extend the methodology to estimate non-negative variance components \( \boldsymbol{\theta}\) in the longitudinal context, which further provides estimates for heritability at a specific time point, as well as for heritability of velocity in a longitudinal trait, denoted by \(\hat{\lambda}_1\) and \( \hat{\lambda}_2\) constrained between 0 and 1. To simplify notation, we begin by assuming there are no fixed effects. In Supplemental Material B, we show that 
\begin{equation} \label{Eq:General1}
\begin{split}
& \mathbb{E}(y_{ij}^{2} | \boldsymbol{Z}, \boldsymbol{\theta})=  \sigma^2_{g}G_{ii} + \sigma^2_{g^{\ast}} G_{ii} t_{ij}^2 + \sigma^2_{b0} + \sigma^2_{b1}t_{ij}^{2} + \sigma^2_{e},  \\
& \mathbb{E}(y_{ij}y_{ir} | \boldsymbol{Z}, \boldsymbol{\theta})=  \sigma^2_{g}G_{ii} + \sigma^2_{g^{\ast}}G_{ii} t_{ij} t_{ir} + \sigma^2_{b0} + \sigma^2_{b1}t_{ij}t_{ir}, \\ 
& \mathbb{E}(y_{ij}y_{km} | \boldsymbol{Z}, \boldsymbol{\theta})=   \sigma^2_{g} G_{ik} + \sigma^2_{g^{\ast}} G_{ik} t_{ij} t_{km},
\end{split}
\end{equation}
where \(i,k=1,2,\ldots,N\), \(j,r=1,2,\ldots,n_i\), \(m =1,2,\ldots, n_k\), \(j \neq r\) and \(i\neq k\). 

Thus, we estimate parameters by minimizing the loss function 
\begin{equation} \label{Eq:Loss_Function}
    F(\boldsymbol{\theta}) =  \sum_{i=1}^{N} \sum_{j=1}^{n_i} \sum_{h=1}^{N}  \sum_{l=1}^{n_h}  [{y}_{ij} {y}_{hl} - \mathbb{E}({y}_{ij}{y}_{hl}| \boldsymbol{Z}, \boldsymbol{\theta} )]^2, 
\end{equation}
subject to the constraints that all variance components are non-negative. The confidence intervals for \(\hat{\boldsymbol{\theta}} \) and \((\hat{{\lambda}}_1, \hat{{\lambda}}_2 )\) can be obtained by parametric bootstrapping.

When the number of subjects and observations is very large, computing the loss function \(F(\boldsymbol{\theta})\) becomes computationally intensive. To address this challenge, we reformulate the loss function and introduce a fast algorithm that substantially reduces the computational cost.

From Equation~\eqref{Eq:General1} and ~\eqref{Eq:Loss_Function}, the loss function \( F(\boldsymbol{\theta})\) can be expressed as: 
\begin{equation}  \label{Eq: optim_function1}
 F(\boldsymbol{\theta})  = \sum_{i=1}^{N} \sum_{j=1}^{n_i} \Delta_{ij}^2 + 2 \sum_{i \in \mathcal {F}}\sum_{j=1}^{n_i -1} \sum_{r=j+1 }^{n_i} \Delta_{ijr}^2  + 2 \sum_{i=1}^{N-1} \sum_{j=1}^{n_i} \sum_{k= i+1}^{N} \sum_{m=1}^{n_k} \Delta_{ijkm}^2, 
\end{equation}
where \(\mathcal{F}\) denotes the set of indices corresponding to individuals with at least two repeated measurements, and 
\begin{equation*}
    \begin{split}
        & \Delta_{ij} = {y}_{ij}^2 - (\sigma^2_g G_{ii} + \sigma^2_{g^{*}} G_{ii} t_{ij}^2 + \sigma^2_{b0} + \sigma^2_{b1}t_{ij}^2 + \sigma^2_e), \\
        &  \Delta_{ijr} = {y}_{ij}{y}_{ir} - ( \sigma^2_{g} G_{ii} +  \sigma^2_{g^{\ast}}  G_{ii} t_{ij} t_{ir} + \sigma^2_{b0} + \sigma^2_{b1}t_{ij}t_{ir}) , \\
        & \Delta_{ijkm} = {y}_{ij}{y}_{km} - ( \sigma^2_{g} G_{ik} + \sigma^2_{g^{\ast}} G_{ik} t_{ij} t_{km}). 
    \end{split}
\end{equation*}
Moreover, the loss function can be also expressed as a standard convex quadratic function using the matrix form: 
\begin{equation} \label{Eq: KKT}
    F(\boldsymbol{\theta}) = \frac{1}{2} \boldsymbol{\theta}^{\top}\boldsymbol{D} \boldsymbol{\theta} - \boldsymbol{c}^{\top} \boldsymbol{\theta} + \text{constant}, 
\end{equation}
where \(\boldsymbol{D}\) is a \(5\times 5\) constant matrix,  \(\boldsymbol{c}\) is a \(5\times1\) constant column vector,  and both of them solely need to be calculated once (Supplementary Material B). Notably, in the optimization process, \(F(\boldsymbol{\theta})\) is a standard convex quadratic function subject to constraints \( \theta_s \geq 0 \) for all \(s=1,2, 3,4,5\). Then, the optimization process yields a unique solution, making the local minimum also the global minimum, thereby facilitating the estimation of $\boldsymbol{\theta}$ and improving computational efficiency \citep{goldfarb1983numerically,lange2012karush, gordon2012karush}.

When fixed-effect coefficients \(\boldsymbol{\beta}\) are present with design matrix \(\boldsymbol{A}\), we first estimate \(\boldsymbol{\beta}\) using ordinary  least squares: \(\hat{\boldsymbol{\beta}} = (\boldsymbol{A}^{\top}\boldsymbol{A})^{-1}\boldsymbol{A}^{\top}\boldsymbol{y}\). Subsequently, we calculate residues \(\boldsymbol{y}^{\ast} =  \boldsymbol{y} - \boldsymbol{A}\hat{\boldsymbol{\beta}}\) and then adopt the proposed REHE method to estimate \(\boldsymbol{\theta}\) based on residues \(\boldsymbol{y}^{\ast}\).

Using the variance component estimates obtained via the REHE regression method in longitudinal settings, we employ a parametric bootstrapping approach with \(1,000\) repetitions to derive empirical standard errors for the estimated variance components \( \hat{\boldsymbol{\theta}}\) and and the two heritability metrics \(\hat{\lambda}_1\) and \(\hat{\lambda}_2\).  This is achieved by setting the fixed-effect coefficients \(\boldsymbol{\beta}\) and variance components \( \boldsymbol{\theta} \) in Model~\eqref{Eq:model2} to their respective estimated values, \(\hat{\boldsymbol{\beta}}\) and \(\hat{\boldsymbol{\theta}}\).

\section{Simulation Studies} \label{section3_2} 
We conducted extensive simulation studies to evaluate the performance of the proposed AI-REML algorithm and the REHE method, with particular focus on assessing bias and precision for estimating two heritability metrics ($\lambda_1$ and $\lambda_2$). Specifically, we examined the influence of several key factors, including sample size, the number of longitudinal measurements per subject, and the inclusion of non-causal genetic common variants (i.e., $10,000$ causal variants and $6,938,674$ variants unrelated to the phenotype). To make the simulations realistic and tailored for the PSA analysis in the PLCO study, we used the genotype data from the PLCO study to simulate genetic effects $g_i$ for the baseline and $g_i^*$ for the velocity. 

The simulation is done in the following steps. First, we randomly selected \(10,000\) genetic variants as causal. Second, for each subject, we generated independent random genotypic effects $\alpha_p$ for the baseline and $\eta_p$ for the velocity for the \(10,000\) causal variants, and calculated $g_i$ and $g_i^*$ accordingly. After simulating $g_i$ and $g_i^*$, we simulated the ages of selection for each subject and generated the corresponding phenotypic values in Model~\eqref{Eq:psa}. Since the study entry age ranged from 54 to 74 years in the PLCO study and each subject had $J$ repeated measurements in the simulation settings where $J$ was set to be 6 or 10, the screening ages $t_{ij}$ were simulated within the range of 54 to 84 and normalized to $(t_{ij}-54)/30$ to ensure numerical stability.  Coefficients of fixed effects were set as \((\beta_0, \beta_1)^{\top} =  (-0.2118, 0.8415)^{\top}\), similar to the estimates from the PSA analysis in the PLCO study.

Table~\ref{tab:settings_sim} provides parameter settings for Scenarios I, II and III, including the variance components and the corresponding heritability metrics for the intercept and velocity effects. In Scenario I, the residual error is substantially smaller than all other sources of variation including for both genetic ($\sigma_g^2$ and $\sigma^2_{g^{\ast}}$) and extraneous individual-specific ($\sigma^2_{b0}$ and $\sigma^2_{b1}$) components. In this scenario, all non-residual variance components are equal and  both heritability metrics ($\lambda_1$ and $\lambda_2$) are 0.5.  Scenario II  emphasizes stronger heritability of the phenotype at 54 years old ($\lambda_1 = 0.8$)  and weaker heritability for the velocity ($\lambda_2 = 0.2$), while Scenario III reverses this pattern to favor higher heritability of the phenotypic velocity ($\lambda_2 = 0.8$). These settings were chosen to evaluate the robustness and sensitivity of the AI-REML algorithm and the REHE method under situations where the heritability metrics are centered at 0.5, and near the boundary values of 0 or 1. To assess the impact of those aforementioned key factors,  we considered a range of simulation settings with either $N=2,000$ or $15,260$ subjects, each having $J$ longitudinal observations where $J$ was set to be 6 or 10.  The number of genetic variants used to estimate variance components and the two heritability metrics were set to be either $P=10,000$ (causal variants only) or $P=6,948,674$ (including both causal and noncausal variants).  Each simulation setting was evaluated based on $1,000$ repetitions. Notably, for the AI-REML algorithm, simulations were conducted for $N=2,000$, where partitioning was not required and for $N=15,260$ where the datasets were evenly divided into seven subsamples to accommodate computational constraints.

Simulation results are presented in Tables~\ref{tab:ratio1_performance_MAD_KKT} and \ref{tab:ratio2_performance_MAD_KKT}, including both the mean and median of the estimates across simulated realizations. For the proposed AI-REML algorithm using truncation-adjusted meta-analysis in large-scale studies ($N=15,260$), the estimation of both two heritability metrics is unbiased under all situations considered. Interestingly, estimates of both heritability metrics remain unbiased even when all available variants ($P=6,948,674$), rather than all true causal variants,  are included in the model, where the Gaussian effect size assumption for $\alpha_p$ and $\eta_p$ is violated (only $10,000$ causal variant effects simulated from Gaussian distributions and the remaining $6,938,674$ non-causal variant effects set to zero). These findings are consistent with previous simulation studies in cross-sectional settings \citep{yang2010common, yang2011gcta} and are further supported by theoretical results from asymptotic analysis \citep{jiang2016high}. Furthermore, we compared the results of truncation-adjusted meta-analysis (Tables~\ref{tab:ratio1_performance_MAD_KKT} and \ref{tab:ratio2_performance_MAD_KKT}) with those of the simple averaging method (Table S5 of Supplementary Material D) for combining the partition-specific estimates, using the AI-REML algorithm for $N=15,260$. For estimating $\lambda_1$, both methods yielded nearly unbiased estimates. This is because the partition-specific estimates of $\lambda_1$ are typically well-estimated when the sample size per subgroup is large enough, resulting in rarely approaching either the boundary values of 0 or 1. In contrast, the estimation of $\lambda_2$ exhibits substantially greater variability under limited number of measurements per subject, resulting in partition-specific estimates more frequently hitting the boundaries. When $\lambda_2$ is $0.5$, both methods remain nearly unbiased due to the symmetric distribution and equal likelihood of reaching the lower or upper bound. However, when $\lambda_2$ deviates from 0.5, the simple averaging method performed poorly for estimating $\lambda_2$, particularly under the cases with $P=6,948,674$ and $J=6$ (Table S5 of Supplementary Material D), highlighting the necessity of truncation-adjusted meta-analysis in such cases.  As shown in Figure S1 of Supplementary Material D, these cases corresponds to the situation where partition-specific estimates of $\lambda_2$ have a high likelihood of hitting one of the boundary values (0 or 1), with a different (and asymmetric) probability of reaching each boundary.  In contrast, when the number of repeated measures is increased to $J=10$, or when only the causal variants are included ($P=10,000$), the simple averaging method performs well for estimating $\lambda_2$ (Table S5 of Supplementary Material D), as the variability of the partition-specific estimates is reduced, leading to these estimates rarely reaching the boundary values. 

For the AI-REML algorithm with a smaller sample size ($N=2,000$), there is no bias for estimating $\lambda_1$, while estimation of $\lambda_2$ appears biased when its true value deviates from 0.5, the number of measurements is small ($J=6$), and all available variants are incorporated ($P=6,948,674$). However, when we define bias with respect to the median rather than the mean, there is no bias in these cases (due to many of the $\lambda_2$ from 1,000 repetitions being either left-truncated at zero or right-truncated at 1).


Unlike AI-REML, which achieves unbiased estimation for two heritability components when $N=15,260$, REHE provides biased estimation of $\lambda_2$ when its true value deviates from 0.5 and non-causal variants are included ($P=6,948,674$). This bias is eliminated or substantially reduced when bias is based on using the median rather than the mean to account for truncation. 


The results indicate that for both sample sizes ($N$) and differing number of follow-up measurements per subject ($J$),  AI-REML achieves smaller variability in estimating heritability components than REHE. 
This finding is expected since AI-REML, without partitioning when $N=2,000$, utilizes the full likelihood information, whereas REHE relies solely on pairwise information. For the larger sample sizes ($N=15,260$), where partitioning is necessary for implementing AI-REML, the fact that the average Hessian-based standard errors (SE), empirical standard errors (Emp SE), and scaled median absolute median (MAD) \citep{rousseeuw1993alternatives} are similar for differing number of $J$ and $P$ demonstrates the good statistical properties of the proposed truncation-adjusted meta-analysis approach. 

\begin{table} 
\caption{Parameter settings used in Scenarios I–III, based on $1{,}000$ repetitions, where $N = 2{,}000$ (or $15{,}260$), $J = 6$ (or $10$), and $P = 10{,}000$ (or $6{,}948{,}674$). \label{tab:settings_sim}}
\begin{center}
\begin{tabular}{lccc}
\hline
\textbf{Parameter} & \textbf{Scenario I} & \textbf{Scenario II} & \textbf{Scenario III} \\
\hline
$\sigma^2_{g}$     & 2.0   & 2.0   & 0.5 \\
$\sigma^2_{g^*}$   & 2.0   & 0.5   & 2.0 \\
$\sigma^2_{b0}$    & 2.0   & 0.5   & 2.0 \\
$\sigma^2_{b1}$    & 2.0   & 2.0   & 0.5 \\
$\sigma^2_{e}$     & 0.1   & 0.1   & 0.1 \\
$\lambda_1$        & 0.5   & 0.8   & 0.2 \\
$\lambda_2$        & 0.5   & 0.2   & 0.8 \\
\hline
\end{tabular}
\end{center}
\end{table}

\begin{sidewaystable} 
    \centering
    \caption{Comparison of Two Approaches for Heritability on Intercept}
    \label{tab:ratio1_performance_MAD_KKT}
    \scriptsize  
    \begin{threeparttable}  
    \begin{tabular}{r|l|c|c|cccc}
        \toprule
        & & & \multicolumn{4}{c}{\textbf{Mean, Median(SE,Emp SE,MAD)}}  \ML
        & & & \multicolumn{4}{c}{\textbf{$N=2,000$}}  \ML
        \textbf{Parameter} & \textbf{Scenario} & \textbf{True} & \textbf{Method} 
        & \textbf{$P=10,000, J=6$} & \textbf{$P=10,000, J=10$} 
        & \textbf{$P=6,948,674, J=6$} & \textbf{$P=6,948,674, J=10$}  \\ 
        \midrule
        \multirow{3}{*}{\( \lambda_1\)} & \textbf{I} & 0.50 & AI-REML &  0.50,0.50(0.07,0.07,0.08)  & 0.50,0.50(0.07,0.07,0.07) & 0.50,0.50(0.19,0.19,0.19) &  0.50,0.51(0.18,0.18,0.18)  \\
        & \textbf{II} & 0.80 &  ~  & 0.80,0.80(0.07,0.07,0.07) & 0.80,0.80(0.06,0.06,0.06)  & 0.79,0.81(0.19,0.17,0.20)  &  0.79,0.81(0.18,0.16,0.18)  \\
        & \textbf{III} & 0.20 &  ~  & 0.20,0.20(0.08,0.08,0.08) & 0.20,0.20(0.07,0.07,0.08) & 0.21,0.20(0.19,0.17,0.19)  &  0.21,0.21(0.18,0.16,0.19) \\
        \midrule
        \multirow{3}{*}{\(\lambda_1\)} & \textbf{I} & 0.50 & REHE  & 0.50,0.49(---,0.11,0.11)  & 0.50,0.49(---,0.12,0.12) & 0.49,0.49(---,0.23,0.23)  &  0.49,0.48(---,0.24,0.25)  \\ 
         & \textbf{II}  & 0.80 &  ~  & 0.77,0.77(---,0.11,0.11)  & 0.77,0.77(---,0.12,0.11)  & 0.73,0.75(---,0.21,0.24)  &   0.72,0.73(---,0.22,0.26)  \\ 
        & \textbf{III} & 0.20 &  ~  & 0.22,0.23(---,0.11,0.11) & 0.23,0.23(---,0.11,0.12)   & 0.26,0.24(---,0.19,0.22)  & 0.27,0.25(---,0.21,0.24)   \\
        \midrule
        & & & \multicolumn{4}{c}{\textbf{$N=15,260$}}  \ML
        \multirow{3}{*}{\( \lambda_1\)}  & \textbf{I} & 0.50 & AI-REML & 0.50,0.50(0.03,0.03,0.03) & 0.50,0.50(0.02,0.02,0.02)	  & 0.50,0.50(0.07,0.07,0.07) &  0.50,0.50(0.06,0.06,0.07) \\
        & \textbf{II} & 0.80 &  ~  & 0.80,0.80(0.02,0.02,0.02)	& 0.80,0.80(0.02,0.02,0.02) & 0.79,0.79(0.07,0.06,0.07)  & 0.79,0.80(0.06,0.06,0.06)		 \\
        & \textbf{III} & 0.20 &  ~  & 0.20,0.20(0.03,0.03,0.03) & 0.20,0.20(0.03,0.03,0.03) & 0.21,0.21(0.07,0.06,0.06)  & 0.21,0.21(0.06,0.06,0.06)	 \\
        \midrule
        \multirow{3}{*}{\( \lambda_1\)} & \textbf{I} & 0.50 & REHE & 0.50,0.50(---,0.03,0.03) & 0.50,0.50(---,0.03,0.03) & 0.51,0.49(---,0.10,0.08) & 0.51,0.49(---,0.10,0.09)  \\ 
        & \textbf{II}  & 0.80 & ~ & 0.80,0.79(---,0.04,0.03) & 0.80,0.80(---,0.04,0.03) & 0.79,0.77(---,0.10,0.09) &  0.79,0.77(---,0.10,0.10) \\ 
         & \textbf{III} & 0.20 &  ~  & 0.20,0.20(---,0.02,0.02) & 0.20,0.20(---,0.03,0.03) & 0.21,0.20(---,0.06,0.06) & 0.21,0.20(---,0.07,0.06) \\
        \bottomrule
    \end{tabular}
   \caption*{\small{Mean is the average of estimates from 1,000 repetitions; Median is the median of estimates from 1,000 repetitions; SE is the estimated standard error derived from estimated Fisher's information; Empirical SE is calculated across repetitions; MAD represents the scaled median absolute deviation so that it corresponds to the SE under Gaussian distribution and  provides a robust measure of variability; A dash (\text{---}) indicates that SE was not available for the REHE method}}
    \end{threeparttable}
\end{sidewaystable} 


\begin{sidewaystable}  
    \centering
    \caption{Comparison of Two Approaches for Heritability on Velocity}
    \label{tab:ratio2_performance_MAD_KKT}
    \scriptsize  
    \begin{threeparttable}  
    \begin{tabular}{r|l|c|c|cccc}  
        \toprule
        & & & \multicolumn{4}{c}{\textbf{Mean, Median(SE,Emp SE,MAD)}}  \ML
        & & & \multicolumn{4}{c}{\textbf{$N=2,000$}}  \ML 
        \textbf{Parameter} &  \textbf{Scenario} & \textbf{True} & \textbf{Method} 
        & \textbf{$P=10,000, J=6$} & \textbf{$P=10,000, J=10$}   & \textbf{$P=6,948,674, J=6$} & \textbf{$P=6,948,674, J=10$}  \\  
        \midrule
        \multirow{3}{*}{\( \lambda_2\)} &  \textbf{I} & 0.50 & AI-REML  & 0.50,0.50(0.13,0.13,0.13) & 0.50,0.50(0.08,0.08,0.08) & 0.51,0.51(0.33,0.30,0.35) & 0.51,0.50(0.21,0.20,0.20) \\
       & \textbf{II} & 0.20 & ~  & 0.21,0.20(0.16,0.15,0.17) & 0.20,0.20(0.09,0.09,0.09) & 0.28,0.19(0.39,0.29,0.28) &  0.23,0.22(0.23,0.19,0.23) \\
       & \textbf{III} & 0.80 & ~ & 0.80,0.80(0.16,0.15,0.17) & 0.80,0.81(0.08,0.08,0.08) & 0.73,0.82(0.39,0.29,0.28)  & 0.78,0.81(0.23,0.20,0.23)  \\
        \midrule
        \multirow{3}{*}{\( \lambda_2\)} & \textbf{I} & 0.50 & REHE  & 0.49,0.49(---,0.39,0.67) & 0.49,0.49(---,0.37,0.53) & 0.50,0.51(---,0.46,0.73) & 0.50,0.51(---,0.44,0.73) \\  
        & \textbf{II} & 0.20 & ~  & 0.33,0.16(---,0.37,0.23) & 0.31,0.19(---,0.34,0.28) & 0.46,0.32(---,0.45,0.47) & 0.45,0.31(---,0.44,0.46) \\  
        & \textbf{III} & 0.80 & ~ & 0.67,0.82(---,0.37,0.27)	& 0.70,0.82(---,0.34,0.26) & 0.54,0.64(---,0.44,0.54) & 0.55,0.63(---,0.43,0.55) \\  
        \midrule
        & & & \multicolumn{4}{c}{\textbf{$N=15,260$}}  \ML
        \multirow{3}{*}{\( \lambda_2\)} & \textbf{I} & 0.50 & AI-REML & 0.50,0.50(0.05,0.05,0.05) & 0.50,0.50(0.03,0.03,0.03) & 0.50,0.50(0.11,0.13,0.12) & 0.50,0.51(0.07,0.08,0.08) \\  
         & \textbf{II} & 0.20 & ~  & 0.20,0.20(0.05,0.06,0.06) & 0.20,0.20(0.03,0.03,0.03) & 0.20,0.20(0.15,0.13,0.14) & 0.20,0.20(0.08,0.08,0.08)	 \\  
        & \textbf{III} & 0.80 & ~  & 0.81,0.81(0.06,0.06,0.06) & 0.80,0.80(0.03,0.03,0.03) & 0.81,0.82(0.14,0.14,0.15)  & 0.81,0.81(0.09,0.09,0.08)  \\  
        \midrule
        \multirow{3}{*}{\( \lambda_2\)} & \textbf{I} & 0.50 & REHE  & 0.50,0.50(---,0.11,0.11) & 0.50,0.50(---,0.09,0.09) & 0.50,0.49(---,0.27,0.28) & 0.50,0.49(---,0.23,0.23)  \\  
        & \textbf{II} & 0.20 & ~  & 0.20,0.20(---,0.13,0.14) & 0.20,0.20(---,0.11,0.11) & 0.27,0.23(---,0.25,0.29) & 0.25,0.22(---,0.22,0.25) \\  
        & \textbf{III} & 0.80 & ~ & 0.80,0.80(---,0.10,0.11) & 0.80,0.80(---,0.09,0.09) & 0.78,0.80(---,0.21,0.26) & 0.79,0.81(---,0.18,0.22) \\  
        \bottomrule
    \end{tabular}
    \caption*{\small{Mean is the average of estimates from 1,000 repetitions; Median is the median of estimates from 1,000 repetitions; SE is the estimated standard error derived from estimated Fisher's information; Empirical SE is calculated across repetitions; MAD represents the scaled median absolute deviation so that it corresponds to the SE under Gaussian distribution and  provides a robust measure of variability; A dash (\text{---}) indicates that SE was not available for the REHE method}}
    \end{threeparttable}
\end{sidewaystable}

\section{Application to PSA Analysis} \label{section4_2}
\begin{figure}[!ht]
\centering
\includegraphics[width=15cm, height=9.27cm]{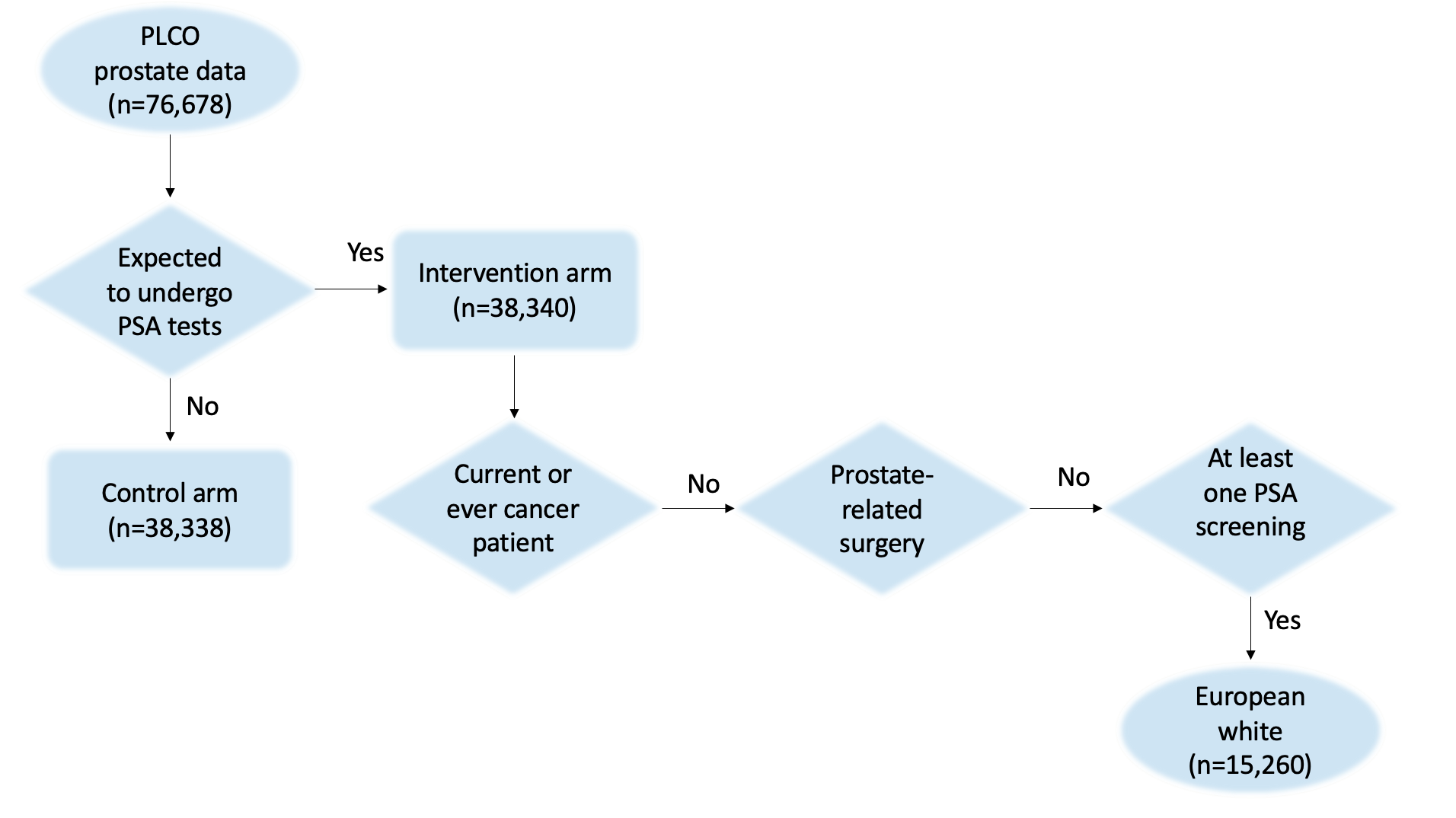}
\caption{Flow chart illustrating the sample selection criterion, detailing each stage of participant inclusion and exclusion, and culminating in the retained sample used for the longitudinal PSA analysis in the PLCO study}
\label{fig:flow_chart}
\end{figure}

With the serial PSA measurements from the Prostate, Lung, Colorectal, and Ovarian (PLCO) Cancer Screening Trial, we can  use our methodologies to simultaneously estimate the heritability for PSA at a particular time point (age) and the velocity among prostate cancer-free participants.  Our analysis focused on the subsample of participants who are European white males in the screening arm,  had no prior history of cancer or prostate-related surgeries (such as transurethral resection of the prostate, prostatectomy, or vasectomy), remained prostate cancer-free throughout the trial,  and had at least one recorded PSA measurement (Figure~\ref{fig:flow_chart}). Participants in this subsample ($N=15,260$) were expected to undergo up to six annual blood tests to measure serum PSA concentrations. The distribution of available records among these participants was as follows: 61.9\% had six records, 21.5\% had five, 9.9\% had four, 3.2\% had three, 2.1\% had two, and 1.4\% had only one record.

This dataset is very unique, as few cohort studies have collected simultaneously genome-wide genetic data in addition to pre-specified serial PSA measurements. For the genetic information, we consider  \( P= 6,948,674\) common variants on autosomal chromosomes,  with a minor allele frequency (MAF) at least \(0.01\) and an imputation quality score \(r^{2} \geq 0.3\) to ensure quality control. In the PLCO genetic data, both the dosage and allele frequency (AF) for each SNP are based on the reference allele. The genetic relationship matrix \(\boldsymbol{G}\) was constructed by the standardized genotype values, defined as \(z_{ip} = (x_{ip} - AF_p)/ \sqrt{2AF_p(1-AF_p)}\) where \(x_{ip} \) is the number of copies of the reference allele for the \(p\)th SNP of the \(i\)th individual, and \(AF_p\) is the frequency of the reference allele for the \(p\)th SNP.

We propose a longitudinal model for the serial PSA measurements  that incorporates screening age and joint genetic effects derived from genome-wide genetic variants to describe the trajectories among the 15,260 unaffected males:
\begin{equation} \label{Eq:psa}
    y_{ij} = \beta_0 + \beta_1 t_{ij} + g_i + g^{*}_i t_{ij} + b_{i0} + b_{i1}t_{ij} + e_{ij}, g_i = \sum_{p=1}^{P} \alpha_p z_{ip}, g_i^* = \sum_{p=1}^{P} \eta_p z_{ip}.
\end{equation}

\begin{figure}[ht]
\centering
\includegraphics[width=12cm, height=7.14cm]{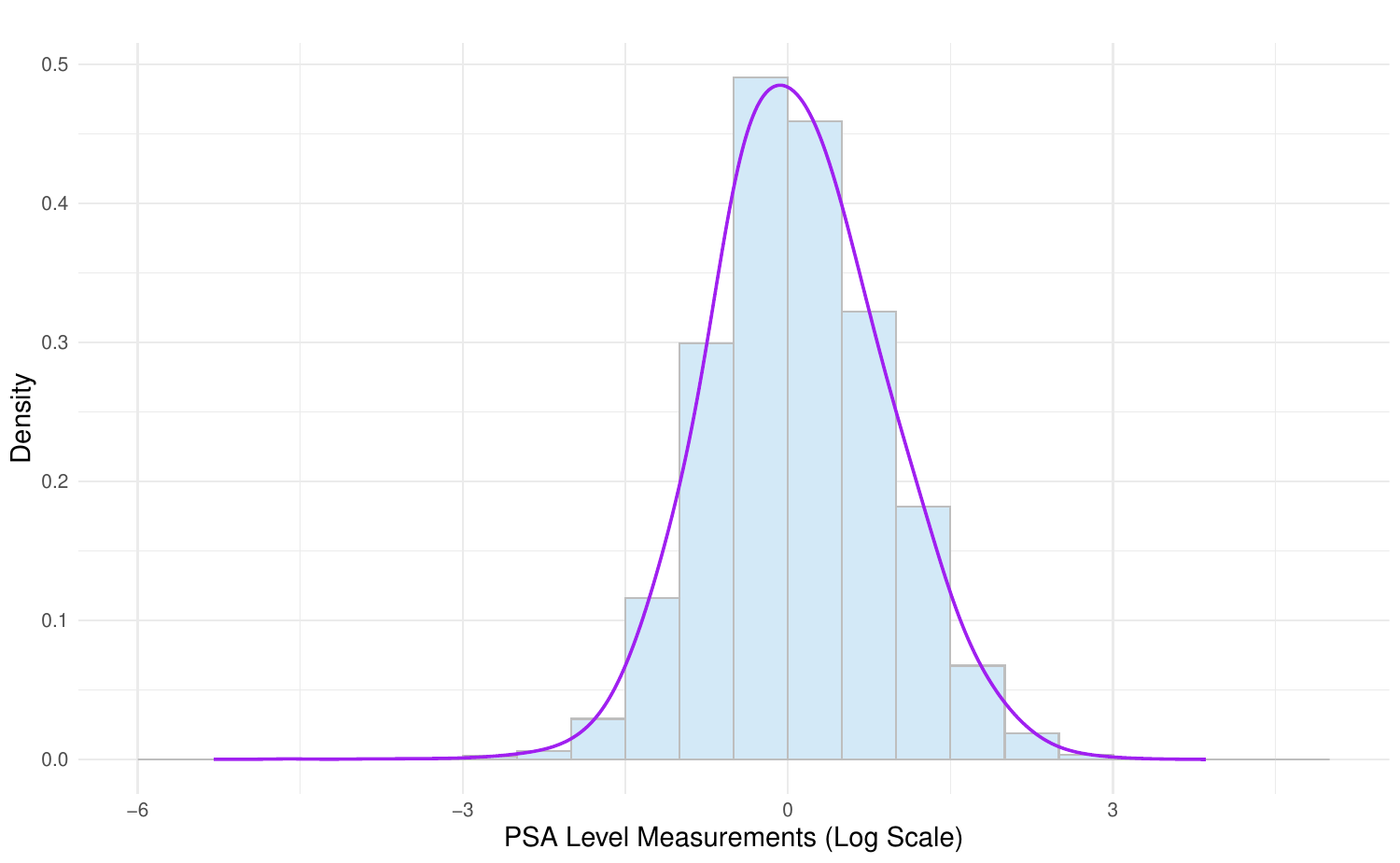}
\caption{Histogram and Empirical Distribution of PSA Measurements on Log Scale Among Retained 15,260 Males}
\label{fig:PSA_log_15260}
\end{figure}

Recognizing the skewness of measurements of PSA level in this population, 
we applied a log transformation to normalize the distribution and meet the normality assumption for the response in Model~\eqref{Eq:psa} \citep{zhang2025mixed}. A small subset of participants (9 out of $15,260$) were recorded zero PSA values. To enable log transformation, these zero values were replaced with $0.005$, as the minimum observed non-zero PSA level prior to transformation was 0.01. Following this adjustment, the empirical distribution of the transformed PSA measurements closely approximates a Gaussian distribution, as demonstrated in Figure~\ref{fig:PSA_log_15260}. We rescaled the time variable $t$ among the male population such that the range from the youngest age at enrollment (54) to the oldest age at the end of follow-up (80) approximately spans 1.

A direct application of AI-REML for fitting Model~\eqref{Eq:psa} would require the inversion of a \(81, 454 \times 81,454\)
dimensional covariance matrix since our dataset has $15,260$ participants with an average of $5.34$ measurements per individual; this makes the estimation computationally infeasible. We randomly partition the population into five groups, each containing \(3, 052\) males, and then apply the AI-REML algorithm on each subsample. Variance components and the two derived heritability metrics are separately estimated for each partition.
Table~\ref{tab:estimates_5parts} presents the partition-specific estimates from the AI-REML algorithm (columns 2-6); note that in some cases, these estimates are at the boundary of the parameter space (this only occurs when estimating the variance components and heritability related to slope). To obtain combined estimates of the variance components ($\boldsymbol{\theta}$) and two heritability metrics ($\lambda_1$ and $\lambda_2$) and their standard errors, we performed truncation-adjusted meta-analysis. For each variance component,  we used left-truncated adjustment since there is only a lower bound at zero and no upper bound. For the heritability metrics ($\lambda_1$ and $\lambda_2$), those of which is bounded by 0 and 1,  we used a doubly truncated adjustment. The results indicate substantial heritability for both the intercept and slope, estimated at 0.31 (SE = 0.07) and 0.45 (SE = 0.18), respectively.  As a sensitivity analysis, we performed the analysis by partitioning the population into seven groups and obtained similar results (Supplementary Material D). As seen in the simulation results, standard errors for estimating slope heritability are substantially larger than those for the intercept, demonstrating the difficulty in making inference on the genetic influences for velocity. In the PLCO Study, we had an average number of follow-up measurements across participants of $5.34$. With additional follow-up measurements, we would certainly decrease the standard error for  $\lambda_2$. In the simulations, we showed that there was a substantial decrease in the standard errors for $\lambda_2$ when we increased the number of follow-up measurements per subject from 6 to 10 (Table~\ref{tab:ratio2_performance_MAD_KKT}). 
Moreover, the two estimated heritability metrics, combined from partition-specific estimates using the simple averaging method or fixed-effect meta-analysis, were comparable to those derived from the truncation-adjusted meta-analysis (Supplementary Material E). This consistency is expected, as our simulation studies indicated minimal bias in estimating $\lambda_2$ when this heritability metric is approximately $0.50$.

\begin{table}[ht]
    \centering
    \caption{Estimation Results for Five Evenly Divided Parts Using the AI-REML Algorithm with Meta-Analysis of Truncated Estimates}
    \label{tab:estimates_5parts}
    \begin{tabular}{r|cccccc}
        \toprule
        & \multicolumn{6}{c}{\textbf{Estimate(SE)}} \\
        \cmidrule{2-7}
        \textbf{Parameter} & \textbf{1st Part} & \textbf{2nd Part} & \textbf{3rd Part} & \textbf{4th Part} & \textbf{5th Part} & \textbf{Combined}  \\
        \midrule
        $ \sigma^2_g$      &  0.15(0.07)  & 0.12(0.07)  & 0.19(0.07)  & 0.18(0.07)  & 0.06(0.07) & 0.14(0.03) \\
        $ {\sigma}^2_{g^*}$ &  0.02(0.33)  & 0.00(0.41)  & 0.85(0.36)  & 0.05(0.33)  & 0.94(0.38)  & 0.32(0.16) \\
        $ {\sigma}^2_{b0}$  &  0.31(0.07)  & 0.32(0.07)  & 0.25(0.06)  & 0.26(0.06)  & 0.39(0.07)  & 0.30(0.03) \\
        $ {\sigma}^2_{b1}$  &  0.84(0.33)  & 1.06(0.41)  & 0.01(0.35)  & 0.73(0.32)  & 0.00(0.36)  & 0.49(0.14) \\
        $ {\sigma}^2_e$     &  0.10(0.00)  & 0.10(0.00)  & 0.10(0.00)  & 0.09(0.00)  & 0.10(0.00)  & 0.10(0.00) \\
        $ \lambda_1$        &  0.32(0.16)  & 0.27(0.16)  & 0.44(0.16)  & 0.41(0.15)  & 0.14(0.15)  & 0.32(0.07) \\
        $ \lambda_2$        &  0.02(0.39)  & 0.00(0.39)  & 0.99(0.40)  & 0.07(0.42)  & 1.00(0.39)  & 0.45(0.18) \\
        \bottomrule
    \end{tabular}
    \begin{flushleft}
        \footnotesize SE denotes the estimated standard error, derived from estimated Fisher information. For the combined estimates based on five parts, meta-analysis was performed for each of variance components ($\sigma^2_g$, $\sigma^2_{g^{\ast}}$, $\sigma^2_{b0}$, $\sigma^2_{b1}$ and $\sigma^2_e$) using left-truncated Gaussian distribution, while heritability metrics ($\lambda_1$ and $\lambda_2$) were analyzed using doubly truncated Gaussian distributions.
    \end{flushleft}
\end{table}

Table~\ref{tab:REHE_PSA} shows the results when we applied the REHE method to the PLCO Study data,  we find that the estimated ratios of genetic contributions to the baseline PSA levels and the rate of change in PSA level are \(0.26 \ (\text{SE} = 0.06)\) and \(0.48 \ (\text{SE} = 0.37)\), respectively.  A comparison of the AI-REML algorithm and REHE method shows similar estimates of heritability metrics and standard errors for heritability on the intercept ($\lambda_1$), but notable differences in the standard errors for velocity heritability ($\lambda_2$).  This observation is consistent with findings from the simulation studies, where AI-REML, particularly when combined with truncation-adjusted meta-analysis, demonstrated superior performance over REHE, especially in estimating the velocity heritability. Overall, based on the AI-REML results for the longitudinal PSA data from the PLCO study, the heritability of baseline PSA was estimated at $0.32 \ ( \text{SE} = 0.07)$ and the heritability of PSA velocity at $0.45 \ ( \text{SE} = 0.18)$, suggesting a substantial genetic contribution to PSA velocity.  This finding highlights the potential for genome-wide association studies of PSA velocity and further genetically adjusted PSA velocity to enhance the efficacy of PSA screening for the early detection of (aggressive) prostate cancer.

\begin{table}[ht]
    \centering
    \caption{Results of PSA Analysis Using the Adapted REHE Method and Parametric Bootstrapping with $1,000$ Repetitions}
    \label{tab:REHE_PSA}
    \begin{tabular}{r|ccc}
        \toprule
        \textbf{Parameter} & \textbf{Estimate} & \textbf{Emp SE} & \textbf{MAD} \\
        \midrule
        ${\sigma}^2_{g}$        & 0.12  & 0.03  & 0.03  \\
        ${\sigma}^2_{g^{\ast}}$ & 0.20  & 0.14  & 0.16 \\
        ${\sigma}^2_{b_{0}}$    & 0.35  & 0.03  & 0.03 \\
        ${\sigma}^2_{b_{1}}$    & 0.22  & 0.13  & 0.15  \\
        ${\sigma}^2_{e}$        & 0.10  & 0.00  & 0.00  \\
        $\lambda_1$             & 0.26  & 0.06  & 0.06  \\
        $\lambda_2$             & 0.48  & 0.31  & 0.37  \\
        \bottomrule
    \end{tabular}
    \begin{flushleft}
        \footnotesize Emp SE refers to the empirical standard error; MAD represents the scaled median absolute deviation, a robust alternative to the sample standard deviation, especially in the presence of outliers or non-normal data.
    \end{flushleft}
\end{table}

\section{Discussion}  \label{section5_2}
In this paper, we introduce a mixed modeling framework designed to assess the genetic contributions to a longitudinal phenotype. This framework separately estimates the overall genetic contributions to both the intercept and  slope components, enabling an examination of static and dynamic aspects of a phenotypic trait. We proposed two approaches, AI-REML and REHE, that can  accommodate large sample sizes, making them suitable for typical population-based studies. These methods build upon earlier approaches applied to cross-sectional phenotypes where only a single individual-specific genetic effect needs to be incorporated for characterizing the heritability \citep{yang2010common,yang2011gcta}.

For both small ($N=2,000$) and larger sample sizes ($N=15,260$), AI-REML algorithm outperforms REHE method in our simulation studies. This was particularly pronounced in estimating $\lambda_2$ as compared with $\lambda_1$. For smaller sample sizes where AI-REML can be directly applied without partitioning,  the reduced standard errors for AI-REML over REHE is easily explained since the former maximizes the full likelihood while the latter is a method-of-moments estimator. In the PLCO analysis and in simulations with the larger sample size of $N=15,260$,  AI-REML still outperforms REHE in most situations. This is interesting since partitioning is required for AI-REML, and the required meta-analysis approach does not use cross-partition information. REHE, however, uses pairwise information across the entire large dataset. For very large longitudinal studies (larger than the PLCO study),  REHE may outperform AI-REML. This would be consistent with the results of \citet{zhou2017unified} who demonstrated that a moment-based estimation method, HE, outperformed partitioned AI-REML for cross-sectional studies with very large sample sizes. 

Incorporating genetic variation into the rate-of-change in a longitudinal phenotype poses significant methodological challenges. First, applying the AI-REML or REHE to data with large sample sizes is difficult since the dimensionality increases with the total number of measurements and not just the number of individuals.  A direct implementation of AI-REML is prohibitive for large-scale studies since inverting a covariance matrix whose dimension of the total number of measurements is required. In large studies, we proposed a convenient way to conduct the AI-REML algorithm by evenly splitting the population into multiple groups and then combining estimates across these groups. Particularly for estimating the slope heritability ($\lambda_2$), we proposed a truncation adjustment that accounts for group-specific estimates on the boundary. In the motivating example, estimates using this adjustment were similar to estimates obtained by simply averaging estimates (some on the boundary) across groups. Our simulations showed that the simple averaging method performed well when the true value of  $\lambda_2$ is $0.50$, since group specific estimates were as likely to be 0 as to be 1.  In contrast, simulations showed that the naive approach of simply averaging the estimates performed poorly when $\lambda_2$ was closer to either 0 or 1.

Through analyses of PLCO Study data and simulation studies, we found that a large number of longitudinal measurements is needed for estimating the slope heritability well. For example, in our application where the average number of serial measurements per individual is $5.34$, the standard error for $\hat{\lambda}_2$ is substantially larger than that for $\hat{\lambda}_1$ for the AI-REML algorithm (Table~\ref{tab:estimates_5parts}). In fact, a 95\% confidence interval for $\hat{\lambda}_2$ ranges from $0.09$ to $0.80$, making precise inference on $\lambda_2$ difficult in this analysis. Simulation results suggested that there is substantially reduced variability for estimating $\lambda_2$ when $J=10$ as compared with $J=6$.   

With the proposed methodologies, we are able to estimate the heritability of both a longitudinal biomarker at a particular time point and the velocity of the phenotype. Our method allows for the estimation of heritability at any age ($t$). However, given our formulation, which decomposes overall genetic effects into baseline and velocity components, the resulting time-dependent heritability follows a quadratic pattern with time. If obtaining a time-dependent assessment of heritability is the purpose, a more flexible approach could be developed where non-parametric regression methods  tailored to high-dimensional data structures may be applied \citep{hastie1993varying, csenturk2008generalized, hoover1998nonparametric}. Specifically, instead of modeling two random genetic effects ($g_i$ and $g_i^*$), we might directly model a time-dependent genetic effects as $g_i(t)=\sum_{p=1}^{P}\alpha_p(t)z_{ip}$, where the high-dimensional random functions $\alpha_p(t)$'s follow a stochastic processes with mean zero. This is an area for future research.

We conclude that the proposed AI-REML algorithm combined with truncation-adjusted meta-analysis perform well in estimating the heritability of static and dynamic aspects of longitudinal phenotypes in large population studies. Several alternative REML-based approaches that may offer improved efficiency warrant exploration in future research. First, directly maximizing the restricted log likelihood without partitioning would be a possible if we could invert very high dimensional covariance matrices corresponding to our model framework, where all measurements are correlated with each other (within and between individuals). However, this poses a significant numerical challenge. Second, a straightforward two-stage approach could be considered, in which individual-specific coefficients for intercepts and slopes are estimated in the first stage, followed by the estimation of velocity heritability based on REML method in the second stage. Although this two-stage approach may be feasible under balanced data structures, it may be problematic in settings with highly imbalanced designs. Third, a Monte Carlo Expectation Maximization (MCEM) REML algorithm could be developed by extending methods previously proposed for heritability analysis in cross-sectional studies \citep{rousseeuw1993alternatives, loh2015contrasting, loh2015efficient}) to the longitudinal setting. The development of MCEM and Bayesian approaches to address this problem remains an important direction for future research.

\section*{Acknowledgments}
We express our gratitude to the participants of the Prostate, Lung, Colorectal, and Ovarian (PLCO) Cancer Screening Trial and the scientists who contributed to the development of this invaluable resource. This research used PLCO prostate cancer datasets. This work was supported by the National Institutes of Health (NIH) Intramural Research Program and utilized the computational resources of the NIH HPC Biowulf cluster (\url{https://hpc.nih.gov}).

\section*{Code and Data Availability}
R code used for simulation studies and analysis are provided in a walkthrough available on GitHub at \url{https://github.com/BettyzhangPei/Heritability-Velocity}. Access to the PLCO datasets can be requested through the Cancer Data Access System (CDAS) at \url{https://cdas.cancer.gov/}.

\section*{Supplementary Information}
Supplementary Material A details the notations and derivations involved in the matrix calculations.  
Supplementary Material B illustrates detailed deductions for the proposed methods.
Supplementary Material C outlines efficient preprocessing steps for generating and storing joint genetic effects for simulation studies. It also presents a comparison between the simple averaging method and truncation-adjusted meta-analysis -based on the AI-REML algorithm - for combining estimates from several non-overlapping groups in large-sample settings. Supplementary Material D shows the performance of simulation studies on meta-analysis for truncated estimates with limited sample sizes. Finally, Supplementary Material E presents alternative methods for combining partition-specific estimates obtained from the AI-REML algorithm in the PLCO study, estimation results from seven evenly partitioned subgroups using the AI-REML algorithm, and performance metrics of the REHE method through parametric bootstrapping.



\end{document}